\documentclass[onecolumn]{IEEEtran}
\usepackage{times}

\usepackage[top=0.6in, bottom=0.7in, left=0.6in, right=0.6in]{geometry}

\usepackage{amsmath}  
\usepackage{amssymb}  
\usepackage{mathrsfs} 

\usepackage{theorem}  
\usepackage{cite}     
\usepackage{comment}  

\usepackage{upref}
\usepackage{amsfonts}

\usepackage{soul}

\setstcolor{red}

\usepackage{verbatim}

\usepackage[dvipsnames,usenames]{color}
\usepackage{enumerate}

\usepackage{graphicx}
\usepackage{subfigure}

\usepackage[T1]{fontenc}
\usepackage{amsmath}

\usepackage{latexsym}

\usepackage[draft,ulem=normalem]{changes}
\usepackage[all]{xy}

\usepackage[normalem]{ulem}               
\newcommand\redout{\bgroup\markoverwith
{\textcolor{red}{\rule[0.5ex]{2pt}{0.8pt}}}\ULon}

\usepackage[makeroom]{cancel}
\newcommand\Ccancel[2][red]{\renewcommand\CancelColor{\color{#1}}\cancel{#2}}

\newcommand{\be}[1]{\begin{equation}\label{#1}}
\newcommand{\ee}{\end{equation}}

\newcommand{\bc}{\begin{center}}
\newcommand{\ec}{\end{center}}

\newcommand{\cC}{{\cal C}}
\newcommand{\cD}{{\cal D}}

\newcommand{\cO}{{\cal O}}

\newcommand{\bfa}{{\boldsymbol a}}

\newcommand{\bfc}{{\boldsymbol c}}

\newcommand{\bfv}{{\boldsymbol v}}
\newcommand{\bfw}{{\boldsymbol w}}
\newcommand{\bfx}{{\boldsymbol x}}
\newcommand{\bfy}{{\boldsymbol y}}

\renewcommand{\leq}{\leqslant}

\renewcommand{\geq}{\geqslant}

\newcommand{\Cref}[1]{Co\-rol\-la\-ry\,\ref{#1}}

\theoremstyle{plain} \theorembodyfont{\normalfont\slshape}

\newtheorem{thm}{Theorem$\!$}
\newenvironment{theorem}{\begin{thm}\hspace*{-1ex}{\bf.}}{\end{thm}}

\newtheorem{prop}[thm]{Proposition$\!$}

\newtheorem{lem}[thm]{Lemma$\!$}
\newenvironment{lemma}{\begin{lem}\hspace*{-1ex}{\bf.}}{\end{lem}}

\newtheorem{cor}[thm]{Corollary$\!$}

\newtheorem{prob}[thm]{Problem$\!$}

\newtheorem{defi}[thm]{Definition$\!$}

\newtheorem{claims}{Claim$\!$}
\newenvironment{claim}{\begin{claims}\hspace*{-1ex}{\bf .}}{\end{claims}}

\theorembodyfont{\normalfont}

\newtheorem{exam}{Example$\!$}
\newenvironment{example}{\begin{exam}\hspace*{-1ex}{\bf .}}{\end{exam}}

\newtheorem{remrk}{Remark$\!$}

\definecolor{Codecolor}{named}{White}  

\newcommand{\Copen}{\mbox{\{\kern-5.50pt\{}}
\newcommand{\Cclose}{\mbox{\}\kern-5.50pt\}}}
\newcommand{\Cslash}{\mbox{$\backslash\kern-6.02pt\backslash$}}

\begin{document}

\title{Codes Correcting Two Deletions}
\author{
  \IEEEauthorblockN{
    Ryan~Gabrys\IEEEauthorrefmark{1}~and
    Frederic Sala\IEEEauthorrefmark{2}}\\ 
  {\normalsize
    \begin{tabular}{cc}
      \IEEEauthorrefmark{1}Spawar Systems Center~~ &
      \IEEEauthorrefmark{2}Stanford University\\
           ryan.gabrys@navy.mil & fredsala@stanford.edu \\
    \end{tabular}}
    \vspace{0ex}}    
\maketitle

\begin{abstract} In this work, we investigate the problem of constructing codes capable of correcting two deletions. In particular, we construct a code that requires redundancy approximately $8 \log_2 n + \cO(\log_2 \log_2 n)$ bits of redundancy, where $n$ denotes the length of the code. To the best of the authors' knowledge, this represents the best known construction in that it requires the lowest number of redundant bits for a code correcting two deletions. 
\end{abstract}

\section{Introduction}
This paper is concerned with deletion-correcting codes. The problem of creating error-correcting codes that correct one or more deletions (or insertions) has a long history, dating back to the early 1960's \cite{FF62}. The seminal work in this area is by Levenshtein, who showed in \cite{L65} that the Varshamov-Tenengolts asymmetric error-correcting code (introduced in \cite{VT65}) also corrects a single deletion or insertion. For single deletion-correcting codes, Levenshtein introduced a redundancy lower bound of $\log_2(n) - O(1)$ bits, demonstrating that the VT code, which requires at most $\log_2(n)$ redundancy bits, is nearly optimal.


The elegance of the VT construction has inspired many attempts to extend this code to correct multiple deletions. Such an approach is found in \cite{HF02} where the authors introduce a number-theoretic construction that was later shown in \cite{GFFC12} to be capable of correcting two or more deletions. Unfortunately, even for the case of just two deletions, the construction from \cite{HF02} has a rate which does not converge to one. Other constructions for multiple insertion/deletion-correcting codes such as those found in  \cite{PAGFC12},\cite{RS94} rely on $(d,k)$-constrained codes, and consequently, these codes also have rates less than one.

To the best of the authors' knowledge, the best known construction for two deletions (in terms of the minimum redundancy) can be found in the recent work by Brakensiek et al. \cite{BGZ16}. The authors show that it possible to construct a $t$ deletion-correcting code with $c_t \cdot \log_2 n$ bits of redundancy where $c_t =\cO(t^2 \log_2 t)$. The construction from \cite{BGZ16} is for general $t$ and does not report any {specialized constructions for the case where $t$ is small}. However, it {will be shown in the next section that these} methods result in a construction requiring at least $128 \log_2 n$ bits of redundancy for the case of $t=2$.

\begin{figure}\label{fig:red}\center
\includegraphics[scale=.45]{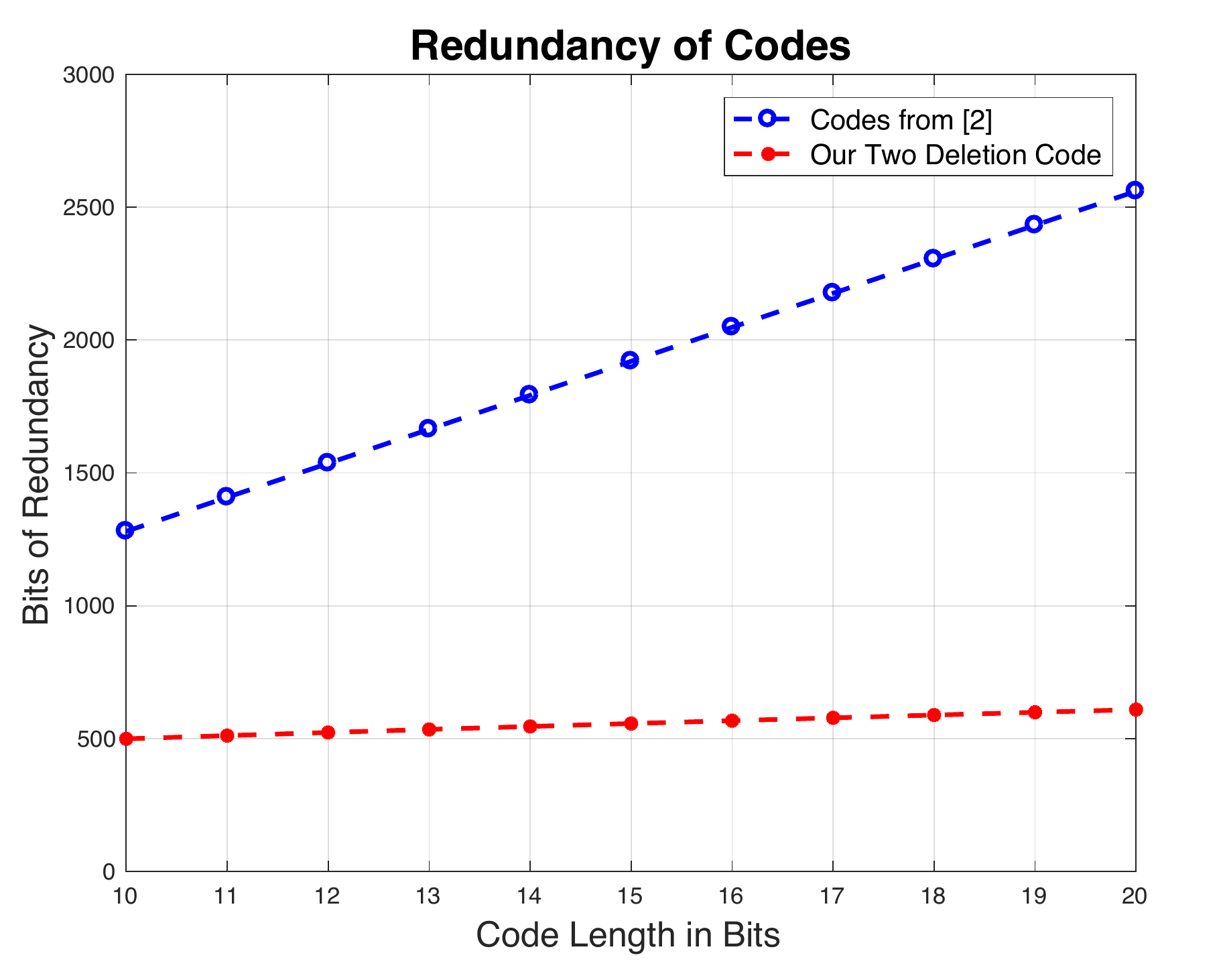}
\caption{Comparison of new codes with codes from \cite{BGZ16}. }
\end{figure}

\begin{figure}\label{fig:rate}\center
\includegraphics[scale=.45]{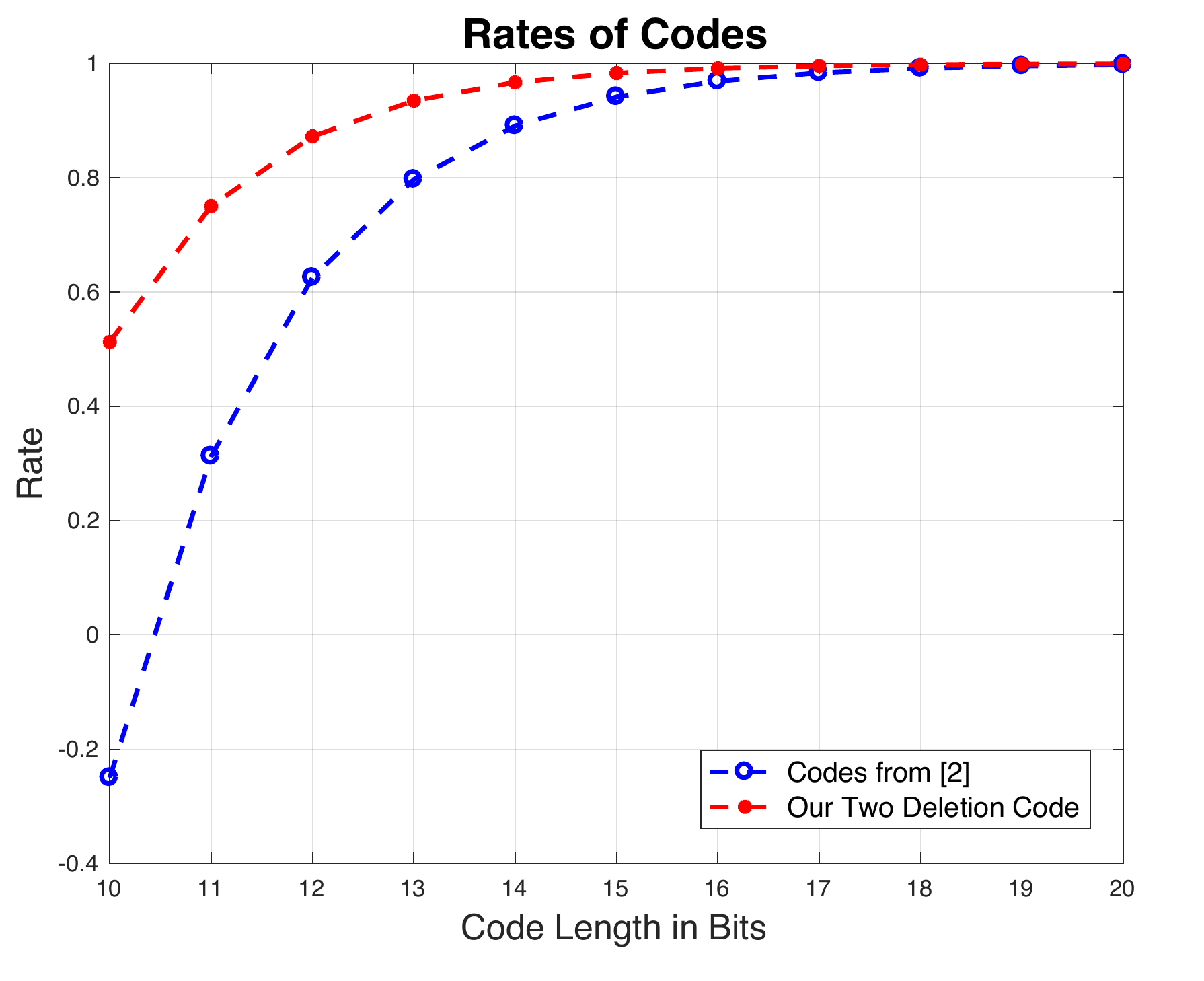}
\caption{Comparison of rate of new codes with codes from \cite{BGZ16}. }
\end{figure}

The best known lower bound for the redundancy of a double deletion-correcting code is $2 \log_2 n - \cO(1)$ (\cite{KK13}) bits; thus, there remains a significant gap between the upper and lower bounds for $t$ deletion-correcting codes even for the case where $t=2$. This motivates the effort to search for more efficient codes.

We note that using a counting argument such as the one found in \cite{L65}, one can show that there exists a $t$ deletion-correcting code with redundancy at most $2t \log _2 n - \cO(t)$. However, these codes require the use of a computer search to form the codebooks along with a lookup table for encoding/decoding. Such codes do not scale as $n$ becomes large and there is no efficient search mechanism that scales sub-exponentially with $n$.

The contribution of the present work is a double deletion-correcting code construction that requires $8 \log_2 n + \cO(\log_2 \log_2 n)$ bits of redundancy. To the best of the authors' knowledge this represents the best construction for a double deletion-correcting code in terms of the redundancy; it is within a factor of four of the optimal redundancy. In Figures~1 and 2, {we compare our construction  and the construction from \cite{BGZ16}, where make use of the expressions (\ref{eq:fredund}), (\ref{eq:qr1}), and (\ref{eq:qr2}), which are derived later in Section~\ref{sec:improved}. To ease the comparison we assume that $\log_2 | \cC_{T2}(n,s)| = n$, and that the number of mixed strings from \cite{BGZ16} of length $n$ is $2^n$.}



The paper is organized as follows. In Section~\ref{sec:main}, we provide the main ideas behind our construction and provide an outline of our approach. Section~\ref{cons1} introduces our first construction. Afterwards, an improved construction is described in Section~\ref{cons2}. We discuss the issue of run-length-limited constrained codes in Section~\ref{sec:constraint} and conclude with Section~\ref{conc}.

\section{Main Ideas and Outline}
\label{sec:main}

The idea behind our approach is to isolate the deletions of zeros and ones into separate sequences of information, and to then use error correction codes on the substrings appearing in the string. We also use a series of constraints that allow us to detect what types of deletions occurred, and consequently we are able to reduce the number of codes in the Hamming metric which are used as part of the construction. As a result, we present a construction which achieves the advertised redundancy.

Let $\cC(n) \in \mathbb{F}_2^n$ denote our codebook of length $n$ that is capable of correcting two deletions. Suppose $\bfy \in \mathbb{F}_2^{n-2}$ is received where $\bfy$ is the result of two deletions occurring to some vector $\bfx \in \cC(n)$, denoted $\bfy \in \cD_2(\bfx)$. Then, we have the following $6$ scenarios:
\begin{enumerate}
\item Scenario 1: Two zeros were deleted from $\bfx$.
\item Scenario 2: Two ones were deleted from $\bfx$.
\item Scenario 3: A zero was deleted from a run of of length $\geq 4$ and a one was deleted from a run of length $\geq 4$ in $\bfx$.
\item Scenario 4: A symbol $b \in \mathbb{F}_2$ was deleted from a run of length $\geq 4$ and another symbol $\bar{b}$ was deleted from a run of length $\ell \in \{1,2,3\}$. Furthermore, if $\ell=1$, then $\bar{b}$ is adjacent to runs of lengths $\ell_1, \ell_2$ where $\ell_1 + \ell_2 < 4$.
\item Scenario 5: A symbol $b \in \mathbb{F}_2$ was deleted from a run of length $\geq 4$, and a symbol $\bar{b}$ is deleted from a run of length $1$, where $\bar{b}$ is adjacent to runs of lengths $\ell_1, \ell_2$ where $\ell_1 + \ell_2 = 4$.
\item Scenario 6: Scenarios 1)-5) do not occur.
\end{enumerate}
The first $5$ of these scenarios are shown in Figure~3.

\begin{figure}\label{fig:5scen}
\centering
\includegraphics[scale=.55]{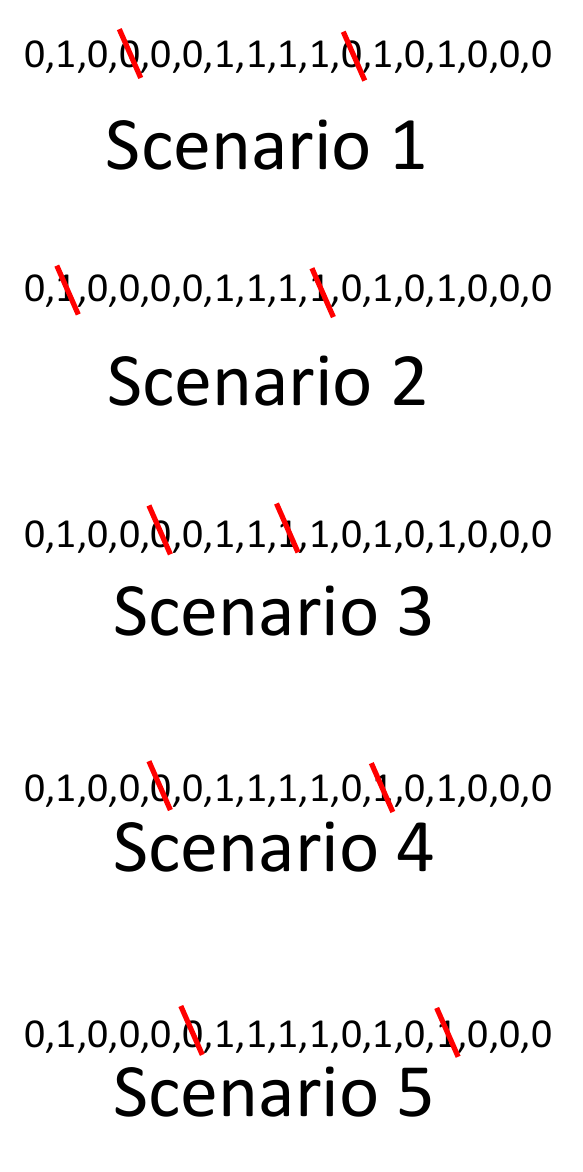}
\caption{First $5$ scenarios}
\end{figure}

Our approach is to use a series of detection codes to attempt to delineate between the $6$ scenarios enumerated above. In addition, and similar to \cite{BGZ16}, we make use of substrings that are not affected by the deletions. The main difference between our approach here and the one used by \cite{BGZ16} is that we use a series of detection codes that allow us to place error-correcting codes on fewer substrings occuring in our codewords. We provide an example which illustrates the basic ideas behind the approach in \cite{BGZ16} and it highlights some subsequent notation used throughout the paper. {In the next example, let $L(\bfx, \bfw)$ be an integer which denotes the maximum number of bits between any two occurrences of the substring $\bfw \in \{0,1\}^{*}$ in $\bfx$, and let $f_s : \{ \emptyset \} \cup \mathbb{F}_2^1 \cup \cdots \cup \mathbb{F}_2^s \to [2^{s+1}-1]$ be an injective mapping where $\emptyset$ denotes the null string.}

\begin{example} Suppose $\bfx = (\textcolor{red}{0},1,1,0,0,0,1,0,\textcolor{red}{1},1,0,0)$ is transmitted. Notice that $L(\bfx, 00) = 4$. For shorthand, let $L(\bfx, 00) = s$. Then, 
$$  F_{f_s}(\bfx, 00) = F_{f_4}(\bfx,00) = ( f_4(0,1,1), f_4(\emptyset), f_4(1,0,1,1), f_4(\emptyset)).$$
Thus, if there are $k$ occurrences of the substring $00$, then the sequence ${F}_{f_4}(\bfx,00)$ has length $k+1$. {To see this, note that $k-1$ symbols in $F_{f_s}(\bfx, 00)$ are created by hashing substrings located between every two consecutive appearances of $00$ in $\bfx$. Note that in this example this corresponds to the symbols $f_4(\emptyset), f_4(1,0,1,1)$. In addition, the first symbol in $F_{f_s}(\bfx, 00)$ is formed by hashing the substring which appears before the first occurrence of $00$ and the last symbol in $F_{f_s}(\bfx, 00)$ is formed by hashing the substring which occurs after the last occurrence of $00$.} 

Suppose $\bfy = (1,1,0,0,0,1,0,1,0,0)$ is received where $\bfy$ is the result of two deletions occurring to $\bfx$. Notice that 
$$ {F}_{f_4}(\bfy, 00) = (f_4(1,1),f_4(\emptyset), f_4(1,0,1), f_4(\emptyset)). $$
In particular, notice that $d_H( {F}_{f_4}(\bfx, 00), {F}_{f_4}(\bfy, 00)) = 2$, where $d_H$ denotes the Hamming distance. Thus, if ${F}_{f_4}(\bfx,00)$ belongs to a double error-correcting code, it is possible to recover ${F}_{f_4}(\bfx, 00)$ from $\bfy$ and in particular, it is possible to then recover $\bfx$. 
\end{example}

Notice in the previous example that all occurrences of the substring $00$ were preserved (we will more rigorously define this notation shortly). It is not too hard to see that if (in the previous example) a deletion occurred where a substring $00$ is deleted in $\bfx$ and therefore does not appear in $\bfy$, then we can no longer claim $d_H( {F}_{f_4}(\bfx, 00), {F}_{f_4}(\bfy, 00)) = 2$. 

In order to overcome this issue, the approach taken in \cite{BGZ16} was to require that the sequence ${F}_{f_4}(\bfx, \bfw )$ belongs to a double error-correcting code for many different choices of $\bfw$. In particular, the approach in \cite{BGZ16} is to enforce that $F_{f_s}(\bfx, \bfw)$ holds for every binary string $\bfw$ of length $m$ where according to Theorem~5 from \cite{BGZ16} we need $2^m >  2t \cdot (2m-1)$. Since any two-error correcting code of length $n$ requires approximately $2 \log_2 n$ bits of redudancy and $t=2$ for our setup, this implies that using the construction from \cite{BGZ16}, we need $m \geq 6$ and so the overall construction requires approximately $2^6 \cdot (2 \log_2 n) = 128 \log_2 n$ bits of redudancy.

The approach taken here is to use a series of detection codes along with different mappings and more carefully choose which substrings to place error-correcting codes on. Consequently, we show it is possible to construct a code with fewer redundant bits than the approach outlined in \cite{BGZ16} for the case of two deletions.

From the above example if we use the constraints ${F}_{f_{s}}(\bfx, \bfw_1)$, ${F}_{f_s}(\bfx, \bfw_2)$, ${F}_{f_{s}}(\bfx, \bfw_3)$ , and ${F}_{f_{s}}(\bfx, \bfw_4)$ (for some appropriately chosen substrings $\bfw_1, \ldots, \bfw_4$), then this would require the use of a series of double error-correcting codes defined over an alphabet of size approximately $2^s$. To reduce the size of this alphabet, we make use of the following lemma.

\begin{lemma}\label{lem:gur} (c.f., \cite{BGZ16}) There is a hash function $h_s : \{0,1\}^s \to \{0,1\}^v$ for $v \leq 4 \log_2 s + O(1)$, such that for all $\bfx \in \{0,1\}^s$, given any $\bfy \in \cD_2(\bfx)$ and $h_s(\bfx)$, the string $\bfx$ can be recovered. \end{lemma}

Codes constructed according to Lemma~\ref{lem:gur} can be found using brute force attempts such as finding an independent set on a graph with $2^n$ vertices for which no polynomial-time algorithms (with respect to $n$) exist. Note, however, that if $s = O(\log_2 n)$, then using an algorithm for determining a maximal independent set on a graph which is polynomial with respect to the number of vertices in the graph (such as \cite{FGK09} for instance) results in a code of length $s$ that can be constructed in polynomial time with respect to $n$.

At first, we will make use of the sequences $F_{h_s}(\bfx,0000)$, $F_{h_s}(\bfx, 1111)$, $F_{h_s}(\bfx, 11011)$, and $F_{h_s}(\bfx, 110011)$ where $\bfx \in \cC(n)$. In particular, we will require that each of these sequences belongs to a code with minimum Hamming distance $5$ over an alphabet of size approximately $s$. Assuming $s = O(\log_2 n)$, then these constraints together require approximately $4 \cdot \frac{7}{3} \log_2(n)$ bits of redundancy if we use the non-binary codes from Dumer \cite{D95}. Afterwards, we alter one of the maps used in conjunction with our Hamming codes and show it is possible construct a code with $8 \log_2 n + O(1)$ bits of redudancy. We now turn to some additional notation before presenting the construction. 

For a vector $\bfx \in \mathbb{F}_2^n$, let $D(i_1, i_2, \bfx) \in \mathbb{F}_2^{n-2}$ be the result of deleting the symbols in $\bfx$ in positions $i_1$ and $i_2$ where $1 \leq i_1 <  i_2 \leq n$. For example if $\bfx = (0,1,0,1,0,0)$, then $D(2,4,\bfx) = (0,0,0,0)$. Using this notation, we have $\cD_2(\bfx) = \{ \bfy : \exists i_1, \exists i_2, \bfy = D(i_1, i_2, \bfx)\}$. 

Let $\bfw \in \{0,1\}^m$.  Suppose $\bfy \in \cD_2(\bfx)$. Then we say that the substring $\bfw \in \{0,1\}^m$ is preserved from $\bfx$ to $\bfy$ if, for every occurrence of $\bfw$, there exists indices $i_1$ and $i_2$ such that $D(i_1, i_2, \bfx) = \bfy$ and the following holds:
\begin{enumerate}
\item $\bfw \not \in $ $\{ (x_{i_1},$ $x_{i_1+1},$ $\ldots,$ $x_{i_1+m-1}),$ $(x_{i_1-1},$ $x_{i_1},$ $\ldots,x_{i_1+m-2}), \\
$\ldots,$ $ $(x_{i_1-m+1},$ $x_{i_1-m+2},  $\ldots$ ,x_{i_1})  \}$,  
\item $\bfw \not \in $ $\{ (x_{i_2},$ $x_{i_2+1},$ $\ldots,$ $x_{i_2+m-1}),$ $(x_{i_2-1},$ $x_{i_2},$ $\ldots,x_{i_2+m-2}), \\
$\ldots,$ $ $(x_{i_2-m+1},$ $x_{i_2-m+2},  $\ldots$ ,x_{i_2})  \}$,
\item $\bfw \not \in $ $\{ (y_{i_1-1},$ $y_{i_1},$ $\ldots,$ $y_{i_1+m-2}),$ $(y_{i_1-2},$ $y_{i_1-1},$ $\ldots,y_{i_1+m-3}), $\ldots,$ $ $(y_{i_1-m+1},$ $y_{i_1-m+2},  $\ldots$ , y_{i_1})  \}$,
\item $\bfw \not \in $ $\{ (y_{i_2-2},$ $y_{i_2-1},$ $\ldots,$ $y_{i_2+m-3}),$ $(y_{i_2-3},$ $y_{i_2-2},$ $\ldots,y_{i_1+m-4}), $\ldots,$ $ $(y_{i_2-m},$ $y_{i_1-m+1},  $\ldots$ , y_{i_2-1})  \}$.
\end{enumerate}
In words, the first two statements above require that any substring $\bfw$ is not deleted from $\bfx$ and the last two statements require that no new appearances of $\bfw$ are in $\bfy$ that were not also in $\bfx$. If $\bfw$ is not preserved from $\bfx$ to $\bfy$, and the first two conditions above are violated, then we say that $\bfw$ was destroyed from $\bfx$ to $\bfy$. If $\bfw$ is not preserved, and the last two conditions above are violated, then we say that $\bfw$ was created from $\bfx$ to $\bfy$. Notice that in order for $\bfw$ to be preserved from $\bfx$ to $\bfy$, 1)-4) has to hold for at least one pair of $i_1, i_2$ such that we can write $\bfy = D(i_1, i_2, \bfx)$ since the choice of $i_1, i_2$ may not be unique. The following example shows this.

\begin{example} Suppose $\bfx = (0,0,1,0,1,\textcolor{red}{1},0,1,0,{0},1,0,1,1,\textcolor{red}{1},0)$ and $\bfy = (0,0,1,0,1,0,1,0,0,1,0,1,1,0)$. Then we say that $(1,0,1)$ is preserved from $\bfx$ to $\bfy$ since there are three occurrences of $101$ in $\bfx$ and $D(6,15,\bfx) = D(5,15,\bfx) = \bfy$. {In particular, the first occurrence of $(1,0,1)$ is preserved since we can write $\bfy = D(6,15,\bfx)$ and the second occurrence of $(1,0,1)$ is preserved since we can write $\bfy = D(5,15,\bfx)$.} Notice that $(1,1,1)$ is not preserved from $\bfx$ to $\bfy$ and in particular $(1,1,1)$ is destroyed. \end{example}

For a vector $\bfx \in \mathbb{F}_2^n$, let $N_0(\bfx)$ denote the number of zeros in $\bfx$. Similarly, let $N_{1}(\bfx)$ be the number of ones that appear in $\bfx$. Furthermore, let {$N_{0000}(\bfx), N_{1111}(\bfx)$, $N_{11011}(\bfx)$ be the number of appearances of the substrings $0000$, $1111$, and $11011$ respectively. We illustrate these notations in the following example.}

\begin{example} Let $\bfx = (0,1,1,1,1,0,0,0,0,0,1,0,1,1,0,1,1$ $,1,0,0,0)$. Then, $N_0(\bfx) = 11$, $N_1(\bfx) = 10$, {$N_{0000}=2$}, {$N_{1111}=1$}, and $N_{11011}(\bfx)=1$. Notice that two occurrences of the substring $0000$ overlap.
\end{example}

{As discussed earlier, in the next section we consider a construction of a code $\cC(n)$ capable of correcting two deletions where for any $\bfx \in \cC(n)$ we have that $F_{h_s}(\bfx,0000)$, $F_{h_s}(\bfx, 1111)$, $F_{h_s}(\bfx, 11011)$, and $F_{h_s}(\bfx, 110011)$ each belong to a code with minimum Hamming distance $5$. We now give some intuition behind the choice of the substrings $0000, 1111, 11011,$ and $110011$. First, the substrings $0000$ and $1111$ were chosen initially because they have large distance between them. More precisely, let $\bfy = \cD_1(\bfx)$, so that $\bfy$ is the result of one deletion occurring to $\bfx$. Then, it can be shown that if the substring $0000$ is not preserved from $\bfx$ to $\bfy$ after one deletion, the substring $1111$ is preserved. Similarly, if the substrings $1111$ and $0000$ are each not preserved, then at least one of the substrings $\{11011, 110011\}$ is preserved (these ideas are formalized in Claims~\ref{cl:if11011} and \ref{cl:2eqs}). Thus, we will show in the next section that it is possible to always recover $\bfx$ provided $F_{h_s}(\bfx,0000)$, $F_{h_s}(\bfx, 1111)$, $F_{h_s}(\bfx, 11011)$, and $F_{h_s}(\bfx, 110011)$ each belong to a code with minimum Hamming distance $5$ (along with some additional constraints) since at least one of the substrings $\{ 0000, 1111, 11011, 110011\}$ is preserved from $\bfx$ to $\bfy$.}

\section{Construction - First Attempt}
\label{cons1}
We now turn to describing out code. Let $\cC_T(n,s)$ denote the following set
\begin{align*}
 \cC_T(n,s) = \{& \bfx \in \mathbb{F}_2^n : L(\bfx,0000) \leq s, L(\bfx,1111)\leq s, L(\bfx, 110011)\leq s, L(\bfx, 11011)\leq s \}.
\end{align*}
As we will see shortly, our main construction will be a sub-code of $\cC_T(n,s)$.

Let {$\bfc \in \mathbb{F}_7^6$}. Suppose $q$ the smallest odd prime greater than the size of the image of the {hash function $h_s$ from Lemma~\ref{lem:gur}}. Suppose $N$ is the smallest positive integer such that $q^{N-1} > n$. Let ${\bfa_{0000}}, {\bfa_{1111}}, \bfa_{110011}, \bfa_{11011} \in \mathbb{F}_q^{r}$ where $r \leq 2N + \lceil \frac{N-1}{3} \rceil$. Our construction is the following:

\begin{align*}
\hspace{-2.5ex}\cC(n,& {\bfa_{0000}}, {\bfa_{1111}}, \bfa_{110011}, \bfa_{11011}, \bfc, s) = \Big \{ \bfx \in \cC_T(n,s) : \\
&N_0(\bfx) \bmod 7 = c_1, N_1(\bfx) \bmod 7= c_2, \\
&{N_{1111}}(\bfx) \bmod 7 = c_3, {N_{0000}}(\bfx) \bmod 7= c_4, \\
&{N_{110011}(\bfx) \bmod 7= c_5, N_{11011}(\bfx) \bmod 7= c_6} \\
&F_{h_s}(\bfx, 0000) \in \cC_2(n,q, {\bfa_{0000}}), \\
&F_{h_s}(\bfx, 1111)  \in \cC_2(n,q, {\bfa_{1111}}), \\
&F_{h_s}(\bfx, 110011) \in \cC_2(n,q,\bfa_{110011}), \\
&F_{h_s}(\bfx, 11011) \in \cC_2(n,q,\bfa_{11011}) \Big \},
\end{align*}
where $\cC_2(n,q,\bfa)$ is a code over $\mathbb{F}_q$ of length $n$. If any of the sequences above that are required to be in codes of length $n$ have lengths $M < n$, then we simply assume the last $n-M$ components of the sequences are equal to zero. 

Let $H$ be a parity check matrix for a double error-correcting code (minimum Hamming distance $5$) from \cite{D95} so that $H \in \mathbb{F}_q^{r \times q^{N-1}}$. We define the double error correcting code $\cC_2(n,q,\bfa)$ so that
\begin{align*}
\cC_2(n,q,\bfa) = \{ \bfx \in \mathbb{F}_q^{q^{N-1}} : H \cdot \bfx = \bfa \}. 
\end{align*}

We now show that given any $\bfy \in \cD_2(\bfx)$, it is possible to recover $\bfx \in \cC(n,{\bfa_{0000}}, {\bfa_{1111}}, \bfa_{110011}, \bfa_{11011}, \bfc, s)$. For shorthand, we refer to $\cC(n,{\bfa_{0000}}, {\bfa_{1111}}, \bfa_{110011}, \bfa_{11011}, \bfc, s)$ as $\cC(n)$. For the remainder of the section, we always assume $\bfx$ is a codeword from $\cC(n)$ and $\bfy \in \cD_2(\bfx)$.

{This section is organized as follows. First, Claims~\ref{cl:if11011} and \ref{cl:2eqs} establish some useful properties regarding the substrings $0000$, $1111$, $11011$, and $110011$. Then, we show that $\bfy$ can be recovered from $\bfx$ by considering the following cases:
\begin{enumerate}
\item Lemmas~\ref{lem:2o} and \ref{lem:zeros} consider Scenarios~1 and 2 from Figure~3 where either two zeros are deleted or two ones are deleted.
\item Lemma~\ref{lem:11} considers a special case of Scenario $3$ where a zero and a one are both deleted from runs of length $4$.
\item Lemma~\ref{lem:finallem} handles instances of Scenarios $4$, $5$, and $6$ where symbols $b,\bar{b}$ are deleted from $\bfx$ and either 1) $b$ is from a run of length $4$ and $\bar{b}$ is adjacent to runs of lengths $\ell_1, \ell_2$ such that $\ell_1 + \ell_2 = 4$ or 2) The substrings $0000,1111$ are preserved from $\bfx$ to $\bfy$.
\item Lemmas~\ref{lem:10}, \ref{lem:01}, and \ref{lem:0111} address the remaining cases.
\end{enumerate}}

The following claims will be used throughout the section. 

\begin{claim}\label{cl:if11011} Suppose a zero is deleted from a run of length one in $\bfx$ and the deletion causes
\begin{enumerate}
\item $11011$ to be created/destroyed,
\item $1111$ to be created,
\end{enumerate}
then the substring $110011$ is preserved from $\bfx$ to $\bfy$. \end{claim}
As an example, the previous claim will be concerned with the following type of deletion:

$$ (\times,\times,\times,1,0,1,1,\Ccancel{0},1,1,\times,\times,\times,\times,\times), $$
where $'\times'$ indicates a symbol which is either a zero or a one and $\Ccancel{0}$ represents a deletion (in this case of a symbol with value $0$).

\begin{IEEEproof} The deletion of a zero from a run of length $1$ can destroy a $11011$ substring only if the middle zero is deleted. In this case, the $110011$ substring is preserved. The deletion of a zero from a run of length $1$ can create a $11011$ substring only if either the first zero is deleted from the substring $11101011$ in $\bfx$ or if the second zero is deleted from the substring $11010111$ in $\bfx$. In either case, the substring $110011$ is preserved from $\bfx$ to $\bfy$. \end{IEEEproof}

\begin{claim}\label{cl:2eqs} Suppose a symbol with value $b \in \mathbb{F}_2$ is deleted from a run of length $\geq 4$ and a symbol with value $b$ is deleted from a run of length $1$ where $${N_{1111}}(\bfx) \neq {N_{1111}}(\bfy), {N_{0000}}(\bfx) \neq {N_{0000}}(\bfy).$$ Under this setup, if $b=1$, the substring $110011$ is preserved. Otherwise, if $b=0$ and $11011$ is not preserved from $\bfx$ to $\bfy$, then $110011$ is preserved.
\end{claim}
For example, the previous claim will be concerned with the following setups. If $b=0$, then one instance of the setup from this claim is

$$ (\times,\times,0,\Ccancel{0},0,0,\times,\times,\times,\times,\times,1,\Ccancel{0},1,\times,\times,\times) $$
and if $b=1$, then another example is
$$ (\times,\times,1,\Ccancel{1},1,1,\times,\times,\times,\times,\times,0,\Ccancel{1},0,\times,\times,\times) .$$
\begin{IEEEproof} Suppose  that a symbol with value $b=0$ is deleted from a run of length $\geq 4$ and another symbol with value $0$ is deleted from a run of length $1$. To begin, notice that the zero which was deleted from the run of length $\geq 4$ cannot create/destroy the substrings $11011$, $110011$, $1111$ from $\bfx$ to $\bfy$. Then, according to Claim~\ref{cl:if11011}, if the deletion of a zero from a run of length $1$ a) creates/destroys the substring $11011$ and b) creates the substring $1111$ from $\bfx$ to $\bfy$, then the substring $110011$ is preserved from $\bfx$ to $\bfy$.

Suppose $b=1$. Under this setup, the substring $110011$ is preserved, and so we can recover $\bfx$ from the constraint $F_{h_s}(\bfx, 110011) \in \cC_2(n,q,\bfa_{110011})$. To see this, we first note that an occurrence of the substring $110011$ is destroyed only if a one is deleted from a run of length $2$ which is not possible under this setup. In addition, an occurrence of the substring $110011$ cannot be created by deleting a one from a run of length $1$ (since this would require that the one is also adjacent to runs of lengths $\ell_1, \ell_2$ with $\ell_1+\ell_2 \geq 4$ since a $0000$ substring is created from $\bfx$ to $\bfy$) or by deleting a one from a run of length $4$. Therefore, $110011$ is preserved from $\bfx$ to $\bfy$ when $b=1$. \end{IEEEproof}\vspace{1.5ex}

We begin with the cases where either $\bfy$ is the result of deleting two zeros or two ones from $\bfx$. The first two lemmas handle Scenarios 1) and 2) from the previous section.
 
\begin{lemma}\label{lem:2o} Suppose $N_1(\bfx) - N_1(\bfy) \bmod 7 = 2$. Then, $\bfx$ can be recovered from $\bfy$. \end{lemma}
For example, the previous claim will be concerned with the following setup:
$$(\times,\times,\times,\times,\Ccancel{1},\times,\times,\times,\times,\times,\times,\times,\times,\Ccancel{1},\times,\times,\times).$$

\begin{IEEEproof} Since $N_1(\bfx) - N_1(\bfy) \bmod 7 = 2$, two ones were deleted from $\bfx$ to obtain $\bfy$. If ${N_{0000}}(\bfx) - {N_{0000}}(\bfy) \equiv 0 \bmod 7$, then $0000$ is preserved since the deletion of a $1$ can create at most $3$ $0000$s and so the deletion of two ones can create at most $6$ $0000$s. Thus, we conclude that {$0000$} is preserved from $\bfx$ to $\bfy$. Since two ones were deleted, clearly no $0000$ substrings were destroyed. Therefore, $0000$ is preserved from $\bfx$ to $\bfy$, and so we can recover $\bfx$ from $\bfy$ using the constraint $F_{h_s}(\bfx, 0000) \in \cC_2(n,q,{\bfa_{0000}})$.

If ${N_{1111}}(\bfx) - {N_{1111}}(\bfy) \equiv 0 \bmod 7$, then $1111$ is preserved, and so we can recover $\bfx$ from $\bfy$ using the constraint $F_{h_s}(\bfx, 1111)  \in \cC_2(n,q, {\bfa_{1111}})$.

Now we assume that both ${N_{1111}}(\bfx) - {N_{1111}}(\bfy) \not \equiv 0 \bmod 7$ and ${N_{0000}}(\bfx) - {N_{0000}}(\bfy) \not \equiv 0 \bmod 7$. Note that this is only possible if a one is deleted from a run of length $\geq 4$ and a one is deleted from a run of length $1$. According to Claim~\ref{cl:2eqs}, we can determine $\bfx$ from  $F_{h_s}(\bfx, 110011) \in \cC_2(n,q,\bfa_{110011})$. \end{IEEEproof}\vspace{1.5ex}

Next we turn to the case where two zeros have been deleted.

\begin{lemma}\label{lem:zeros} Suppose $N_0(\bfx) - N_0(\bfy) \bmod 7 = 2$. Then, $\bfx$ can be recovered from $\bfy$. \end{lemma}
For example, we will be concerned with the following setup:
$$(\times,\times,\times,\times,\Ccancel{0},\times,\times,\times,\times,\times,\times,\times,\times,\Ccancel{0},\times,\times,\times).$$
\begin{IEEEproof} Since $N_0(\bfx) - N_0(\bfy) \bmod 7 = 2$, two zeros were deleted from $\bfx$ to obtain $\bfy$. If ${N_{1111}}(\bfx) - {N_{1111}}(\bfy) \equiv 0 \bmod 7$, then we can recover $\bfx$ from $\bfy$ using the constraint $F_{h_s}(\bfx, 1111) \in \cC_2(n,q, {\bfa_{1111}})$ using the same logic as the previous lemma. In addition, if ${N_{0000}}(\bfx) - {N_{0000}}(\bfy) \equiv 0 \bmod 7$, then $0000$ is preserved, and so we can recover $\bfx$ from $\bfy$ using the constraint $F_{h_s}(\bfx, 0000)  \in \cC_2(n,q, {\bfa_{0000}})$.

Now we assume that both ${N_{1111}}(\bfx) - {N_{1111}}(\bfy) \not \equiv 0 \bmod 7$ and ${N_{0000}}(\bfx) - {N_{0000}}(\bfy) \not \equiv 0 \bmod 7$. Similar to the previous lemma, this is only possible if a zero is deleted from a run of length $\geq 4$ and a zero is deleted from a run of length $1$. We can use the constraint $N_{11011}(\bfx) \bmod 7 = {c_6}$ to determine whether the substring $11011$ is preserved from $\bfx$ to $\bfy$. {Notice that if $11011$ is not preserved from $\bfx$ to $\bfy$, then a $11011$ is destroyed, and the middle zero is deleted. In this case, $N_{11011}(\bfy) \bmod 7 =c_6-1 \bmod 7$, and so we can use the constraint $N_{11011}(\bfx) \bmod 7 = c_6$ to determine whether the substring $11011$ is preserved from $\bfx$ to $\bfy$.} If $11011$ is preserved, then we can determine $\bfx$ from $F_{h_s}(\bfx, 11011) \in \cC_2(n,q,\bfa_{11011})$. If $11011$ is not preserved then we can determine $\bfx$ from  $F_{h_s}(\bfx, 110011) \in \cC_2(n,q,\bfa_{110011})$ according to Claim~\ref{cl:2eqs}. \end{IEEEproof}\vspace{1.5ex}

As a result of the previous two lemmas, we assume in the remainder of this section that $\bfy$ is the result of deleting a symbol with a value $1$ and a symbol with a value $0$. The next $3$ lemmas handle the case where ${N_{0000}}(\bfx) \geq {N_{0000}}(\bfy)$ or ${N_{1111}}(\bfx) \geq {N_{1111}}(\bfy)$. The next lemma covers Scenario 3).

\begin{lemma}\label{lem:11} Suppose ${N_{0000}}(\bfx) - {N_{0000}}(\bfy) \bmod 7 = 1$, and ${N_{1111}}(\bfx) - {N_{1111}}(\bfy) \bmod 7 = 1$. Then, $\bfx$ can be recovered from $\bfy$. \end{lemma}
For example, we will be concerned with the following setup:
$$(\times,\times,0,0,\Ccancel{0},0,\times,\times,\times,\times,\times,\times,1,\Ccancel{1},1,1,\times).$$
\begin{IEEEproof} Since ${N_{0000}}(\bfx) - {N_{0000}}(\bfy) \bmod 7 = 1$ and ${N_{1111}}(\bfx) - {N_{1111}}(\bfy) \bmod 7 = 1$, $\bfy$ is the result of deleting a $1$ and a $0$ from $\bfx$ where both symbols belong to runs of lengths $\geq 4$. Since both symbols were deleted from runs of lengths at least $4$, it follows that no $110011$ substrings were created/destroyed and so we can recover $\bfx$ from $F_{h_s}(\bfx, 110011) \in \cC_2(n,q,\bfa_{110011})$.
\end{IEEEproof}\vspace{1.5ex}

The next two lemmas handle Scenario 4).

\begin{lemma}\label{lem:10} Suppose ${N_{0000}(\bfx)} - {N_{0000}}(\bfy) \bmod 7 = 1$ and ${N_{1111}(\bfx)} - {N_{1111}}(\bfy) \bmod 7 = 0$. Then, $\bfx$ can be recovered from $\bfy$.
\end{lemma}
For example, we will be concerned with the following setup:
$$(\times,\times,0,0,\Ccancel{0},0,\times,\times,\times,\times,\times,\times,0,\Ccancel{1},1,0,\times).$$
\begin{IEEEproof} Since ${N_{0000}}(\bfx) - {N_{0000}}(\bfy) \bmod 7 = 1$, clearly a zero was deleted from a run of zeros of length at least $4$ in $\bfx$. Since ${N_{1111}}(\bfx) - {N_{1111}}(\bfy) \bmod 7 = 0$ and there are exactly two deletions (of a zero and a one), no ones were deleted from runs of ones of length $\geq 4$. Thus, we can recover $\bfx$ from $F_{h_s}(\bfx, 1111)  \in \cC_2(n,q,\bfa_{1111})$.
\end{IEEEproof}

\begin{lemma}\label{lem:01} Suppose ${N_{1111}}(\bfx) - {N_{1111}}(\bfy) \bmod 7 = 1$ and ${N_{0000}}(\bfx) - {N_{0000}}(\bfy) \bmod 7 = 0$. Then, $\bfx$ can be recovered from $\bfy$.
\end{lemma}
For example, the previous lemma is concerned with the following setup:
$$(\times,\times,1,1,\Ccancel{1},1,\times,\times,\times,\times,\times,\times,1,\Ccancel{0},0,1,\times).$$

Finally, we turn to the case where either Scenario 5) or Scenario 6) occurs. The next two lemmas handle the case where either ${N_{0000}}(\bfy) > {N_{0000}}(\bfx)$ or ${N_{1111}}(\bfy) > {N_{1111}}(\bfx)$.

\begin{lemma}\label{lem:0111} Suppose ${N_{0000}}(\bfy) - {N_{0000}}(\bfx) \bmod 7 \in \{1,2,3\}$ or ${N_{1111}}(\bfy) - {N_{1111}}(\bfx) \bmod 7 \in \{1,2,3\}$. Then, $\bfx$ can be recovered from $\bfy$.
\end{lemma}
For example, we will be concerned with the following setup:
$$(\times,0,0,0,\Ccancel{1},0,\times,\times,\times,\times,\times,\times,1,\Ccancel{0},0,1,\times),$$
or
$$(\times,1,1,1,\Ccancel{0},1,\times,\times,\times,\times,\times,\times,0,\Ccancel{1},1,0,\times).$$
\begin{IEEEproof} Suppose first that ${N_{0000}}(\bfy) - {N_{0000}}(\bfx) \bmod 7 \in \{1,2,3\}$. Then, clearly a one from a run of length $1$ was deleted in $\bfx$ resulting in the creation of new $0000$ substrings. If ${N_{1111}}(\bfy) - {N_{1111}}(\bfx) \bmod 7 = 0$, then $1111$ is preserved from $\bfx$ to $\bfy$ and so we can recover $\bfx$. 

Otherwise, if ${N_{0000}}(\bfy) - {N_{0000}}(\bfx) \bmod 7 \in \{1,2,3\}$ and ${N_{1111}}(\bfy) - {N_{1111}}(\bfx) \bmod 7 \in \{1,2,3\}$, it follows that a one was deleted from a run of length $1$ and also a zero was deleted from a run of length $1$, so that we have the following type of setup:
$$(\times,0,0,0,\Ccancel{1},0,\times,\times,\times,\times,\times,1,1,\Ccancel{0},1,1,\times),$$
The deletion of a one from a run of length $1$ cannot or create destroy a $110011$ substring or a $11011$ substring since the $1$ needs to be adjacent to two runs of lengths $\ell_1, \ell_2$ where $\ell_1 + \ell_2 \geq 4$ since at least one $0000$ substring is created from $\bfx$ to $\bfy$. Furthermore, the deletion of the one from a run of length $1$ clearly cannot create a $1111$ substring from $\bfx$ to $\bfy$. Therefore, the deletion of  the zero from a run of length $1$ creates a $1111$ substring from $\bfx$ to $\bfy$ and from Claim~\ref{cl:if11011}, if $11011$ is not preserved from $\bfx$ to $\bfy$, then $110011$ is preserved. Thus, we can use the constraints $N_{11011}(\bfx) \bmod 7 = {c_6}$, $F_{h_s}(\bfx, 110011) \in \cC_2(n,q,\bfa_{110011})$ and $ F_{h_s}(\bfx, 11011) \in \cC_2(n,q,\bfa_{11011}) $ to determine $\bfx$.

Notice that it is not possible to have ${N_{0000}}(\bfy) - {N_{0000}}(\bfx) \bmod 7 \in \{1,2,3\}$ and ${N_{1111}}(\bfx) - {N_{1111}}(\bfy) \bmod 7 \in \{1,2,3\}$. This is because in order to have  ${N_{0000}}(\bfy) - {N_{0000}}(\bfx) \bmod 7 \in \{1,2,3\}$, a one is deleted from a run of length $1$ and this creates a new run of zeros of length at least $4$. Then, the deletion of the zero (which also occurs by assumption) can only create $1111$ substrings from $\bfx$ to $\bfy$ and so  ${N_{1111}}(\bfx) - {N_{1111}}(\bfy) \bmod 7 \not \in \{1,2,3\}$.

The case where ${N_{1111}}(\bfy) - {N_{1111}}(\bfx) \bmod 7 \in \{1,2,3\}$, but ${N_{0000}}(\bfy) - {N_{0000}}(\bfx) \bmod 7 \not \in \{1,2,3\}$ can be handled using the same logic as before.
\end{IEEEproof}\vspace{1.5ex}

We have one case left to consider.

\begin{lemma}\label{lem:finallem} Suppose ${N_{1111}}(\bfx) - {N_{1111}}(\bfy) \bmod 7 = 0$, and ${N_{0000}}(\bfx) - {N_{0000}}(\bfy) \bmod 7 = 0$. Then, $\bfx$ can be recovered from $\bfy$.
\end{lemma}
\begin{IEEEproof}{Using the same logic as before, ${N_{1111}}(\bfx) - N_{1111}(\bfy) \bmod 7 = 0$ and $N_{0000}(\bfx) - N_{0000}(\bfy) \bmod 7 = 0$, there are two setups to consider:}
\begin{enumerate}
\item {A symbol $b$ was deleted from a run of length $\geq 4$ and another symbol $\bar{b}$ from a run of length $1$ was deleted which was adjacent to a run of length $\ell_1$ and another run of length $\ell_	2$ such that $\ell_1 + \ell_2 = 4$.}
\item {The $0000$ and $1111$ substrings were preserved from $\bfx$ to $\bfy$.}
\end{enumerate}
{The decoding procedure is the following. Suppose $N_{110011}(\bfx) \equiv N_{110011}(\bfy) \bmod 7$. Then we estimate $\bfx$ to be the sequence which agrees with at least two of the following three constraints:  $\Big \{ F_{h_s}(\bfx, 0000) \in \cC_2(n,q,\bfa_{0000}),$ $F_{h_s}(\bfx, 1111)  \in \cC_2(n,q,\bfa_{1111}),$  $F_{h_s}(\bfx, 110011) \in \cC_2(n,q,\bfa_{110011}) \Big \}$. Otherwise, if $N_{110011}(\bfx) \not \equiv N_{110011}(\bfy) \bmod 7$, we estimate $\bfx$ to be the sequence which agrees with the constraint $ F_{h_s}(\bfx, 0000) \in \cC_2(n,q,\bfa_{0000})$.}

{First suppose that $N_{110011}(\bfx) \equiv N_{110011}(\bfy) \bmod 7$ and suppose that 1) holds. If $b=0$, then the decoding is correct since in this case $\bfx$ agrees with the constraints $ F_{h_s}(\bfx, 1111)  \in \cC_2(n,q_1,\bfa_2)$ and $F_{h_s}(\bfx, 110011) \in \cC_2(n,q,\bfa_{110011})$, since it is not possible to create a $0000$ substring and also to create/destroy $110011$.  Now, suppose $b=1$. In this case, if $N_{110011}(\bfx)  \equiv N_{110011}(\bfy) \bmod 7$, then the deletion of a zero from a run of length $1$ did not create/destroy the substring $110011$ and so $\bfx$ agrees with at least two of the three constraints from  $\Big \{ F_{h_s}(\bfx, 0000) \in \cC_2(n,q,\bfa_{0000}), F_{h_s}(\bfx, 1111)  \in \cC_2(n,q,\bfa_{1111}),  F_{h_s}(\bfx, 110011) \in \cC_2(n,q,\bfa_{110011}) \Big \}$. Thus, the decoding is correct when $N_{110011}(\bfx) \equiv N_{110011}(\bfy) \bmod 7$ and  1) holds.}

{Next consider the case where $N_{110011}(\bfx) \equiv N_{110011}(\bfy) \bmod 7$ and suppose that 2) holds. Then clearly, $\bfx$ agrees with two of the three constraints $\Big \{ F_{h_s}(\bfx, 0000) \in \cC_2(n,q,\bfa_{0000}), F_{h_s}(\bfx, 1111)  \in \cC_2(n,q,\bfa_{1111}),  F_{h_s}(\bfx, 110011) \in \cC_2(n,q,\bfa_{110011}) \Big \}$ and so the decoding is correct in this case.}

{Suppose now that $N_{110011}(\bfx) \not \equiv N_{110011}(\bfy) \bmod 7$ and that 1) holds. Notice that under this setup, $b \not = 0$, since the deletion of a $0$ from a run of length at least $4$ and the deletion of a $1$ from a run of length $1$ that creates a $0000$ substring cannot create/destroy any substrings $110011$ from $\bfx$ to $\bfy$. If $b=1$ and $N_{110011}(\bfx) \not \equiv N_{110011}(\bfy) \bmod 7$, then a zero was deleted from a run of length one and a one was deleted from a run of length $\geq 4$ so that the substring $0000$ was preserved and so the decoding is correct in this case.}

{Finally, we consider the case where $N_{110011}(\bfx) \not \equiv N_{110011}(\bfy)$ $\bmod 7$ and 2) holds. If 2) holds and the $0000$ substring is preserved then clearly $\bfx$ agrees with the constraint $ F_{h_s}(\bfx, 0000) \in \cC_2(n,q,\bfa_{0000})$, and so the decoding procedure is correct in this case as well.} \end{IEEEproof}

As a consequence of Lemmas~\ref{lem:2o}-\ref{lem:finallem}, we have the following theorem.

\begin{theorem} The code $\cC(n, {\bfa_{0000}}, {\bfa_{1111}}, \bfa_{110011}, \bfa_{11011}, \bfc, s)$ can correct two deletions. \end{theorem}

{Recall that if we assume $s = O(\log_2 n)$, then $\cC(n)$ requires at least $4 \cdot \frac{7}{3} \log_2(n)$ bits of redundancy if we use the non-binary codes from Dumer \cite{D95} as a result of the constraints $F_{h_s}(\bfx, 0000) \in \cC_2(n,q,{\bfa_{0000}})$, $F_{h_s}(\bfx, 1111)  \in \cC_2(n,q, {\bfa_{1111}}),$ $F_{h_s}(\bfx, 110011) \in \cC_2(n,q,\bfa_{110011}),$ and $F_{h_s}(\bfx, 11011) \in \cC_2(n,q,\bfa_{11011})$.} In the next section, we make some modifications to the code discussed in this section and afterwards we discuss the redundancy of the resulting code.

\section{An Improved Construction}\label{sec:improved}
\label{cons2} In this section, we modify the construction in the previous section to obtain a code with redudancy $8 \log_2 n + \cO(\log_2 \log_2 n)$. Our construction uses the same substrings to partition our codewords as in the previous section, but we make use of a different hash function in place of $h_s$ from Lemma~\ref{lem:gur}, denoted $h^{(R)}_{s}$. Consequently we show that we can replace the constraint $ F_{h_s}(\bfx, 11011) \in \cC_2(n,q,\bfa_{11011})$ with the constraint that $ F_{h^{(R)}_{s}}(\bfx, 11011)$ belongs to a code with Hamming distance $3$ (rather than Hamming distance $5$). Our analysis and the subsequent proof will mirror the previous section in light of these modifications. This section is organized as follows. We first describe our code construction in detail and then show it has the advertised redudancy. Afterwards, we prove the code can correct two deletions. 

Let $\cC_{T2}(n,s)$ denote the following set where for a binary vector $\bfv$, $\tau(\bfv)$ is the length of the longest run of of zeroes or ones in $\bfv$, 
\begin{align*}
 \cC_{T2}(n,s) = \{& \bfx \in \mathbb{F}_2^n : L(\bfx,0000) \leq s, L(\bfx,1111)\leq s, \\
&L(\bfx, 110011)\leq s,  L(\bfx, 11011)\leq s,\\
& \tau(\bfx) \leq s \}.
\end{align*}

In the following, for a vector $\bfv \in \mathbb{F}_2^n$, let {$\tau_1(\bfv)$} denote the run-length representation of the runs of ones in {$\bfv$}. For example, if $\bfv=(1,1,0,1,0,1,1,1)$, then ${\tau_1}(\bfv) = (2,1,3)$. Furthermore, let ${\tau_{\geq 2}}$ be the run-length representation of ones in {$\bfv$} with lengths at least $2$. For example, ${\tau_{\geq 2}(\bfv)} = (2,3)$.

Let $Q$ be the smallest prime greater than $s$. We now turn to describing the map $h^{(R)}_{s} : \mathbb{F}_2^s \to \mathbb{F}_{Q}^{2}$. Let $H_{R1} \in \mathbb{F}_{Q}^{2 \times s}$ be the parity check matrix for a code $\cC_L$ with Hamming distance at least $3$  over $\mathbb{F}_{Q}$. For a vector $\bfv \in \{0,1\}^s$ we define $h_s^{(R)}$ as the vector which results by considering the run-length representation (as a vector) of the runs of ones in $\bfv$ of length at least $2$ and multiplying $H_{R1}$ by this vector. We provide an example of this map next.

\begin{example} Suppose $\bfv = (0,1,1,0,0,0,1,0,1,1,0,0)$. Then, the vector representing the runs of ones in $\bfv$ is ${\tau}_1(\bfv) = (2, 1, 2)$. Notice that ${\tau}_1(\bfv)$ has an alphabet size which is equal to the length of the longest run in $\bfv$. Then, $h^{(R)}_s(\bfv) = H_{R1} \cdot {\tau}_{\geq 2} (\bfv) = H_{R1} \cdot  (2, 2)$. 
\end{example}

Let $\bfc \in \mathbb{F}_7^6$. Suppose $q_1$ the smallest odd prime greater than the size of the image of the map $h_s$, and let $q_2$ be the smallest prime greater than the size of the image of the map $h_{s}^{(R)}$. As before, let $N_1$ be the smallest positive integer such that $q_1^{N_1-1} > n$, and suppose $N_2$ is the smallest positive integer such that $q_2^{N_2}-1 > n$. Let ${\bfa_{0000}}, {\bfa_{1111}}, \bfa_{110011} \in \mathbb{F}_{q_1}^{r_1}$ and let $\bfa_{11011} \in \mathbb{F}_{q_2}^{r_2}$ where $r_1 \leq 2N_1 + \lceil \frac{N_1-1}{3} \rceil$ and $r_2 \leq 1 + N_2$. In the following, let $b \in \mathbb{Z}_{s+1}$.

Our construction is the following:

\begin{align*}
\cC^{(2)}(n,& {\bfa_{0000}}, {\bfa_{1111}}, \bfa_{110011}, \bfa_{11011}, \bfc, b, s) = \Big \{ \bfx \in \cC_{T2}(n,s) : \\
&N_0(\bfx) \bmod 7 = c_1, N_1(\bfx) \bmod 7= c_2, \\
&{N_{0000}}(\bfx) \bmod 7 = c_3, {N_{1111}}(\bfx) \bmod 7 = c_4, \\
&N_{110011}(\bfx) \bmod 7= c_5, N_{11011}(\bfx) \bmod 7 = c_6 \\
&F_{h_s}(\bfx, 0000) \in \cC_2(n,q_1,{\bfa_{0000}}), \\
&F_{h_s}(\bfx, 1111)  \in \cC_2(n,q_1, {\bfa_{1111}}), \\
&F_{h_s}(\bfx, 110011) \in \cC_2(n,q_1,\bfa_{110011}), \\
&F_{h^{(R)}_{s}}(\bfx, 11011) \in \cC_1(n,q_2,\bfa_{11011}), \\
&\sum_{i \text{\ odd}} {\tau}_1(\bfx)_i = b \bmod {(}s + 1 {)}\Big \},
\end{align*}
where $\cC_1(n,q_2,\bfa_{11011})$ is either a primitive BCH code with roots $\{1,\alpha\}$ ($\alpha \in \mathbb{F}_{q_2^{N_2}}$ is an element of order $q_2^{N_2}-1$) or a coset of such a code. If any of the sequences above that are required to be in codes of length $n$ have lengths $M < n$, then we simply assume the last $n-M$ components of the sequences are equal to zero. 

Since the parameters ${\bfc}, {{\bfa}_{0000}}, {{\bfa}_{1111}}, \bfa_{110011},\bfa_{11011},$ and $b$ can be chosen arbitrarily, it follows using an averaging argument that there exists a choice of ${\bfc}, {{\bfa}_{0000}}, {{\bfa}_{1111}}, \bfa_{110011},\bfa_{11011},$ and $b$ that gives
\begin{align}
|\cC^{(2)}(n,& {\bfa_{0000}}, {\bfa_{1111}}, \bfa_{110011}, \bfa_{11011}, \bfc, b, s)| \geq  \nonumber \\
&\ \ \ \frac{| \cC_{T2}(n,s)|}{7^6  q_1^{3r_1}  q_2^{r_2} (s + 1)  }. \label{eq:fredund}
\end{align}

Assuming that the image of the map $h_s$ has cardinality $2^{4 \log_2(s)}$ and {$s=1065 \log_2(n)$} then we can approximate $q_1 = ({1065} \log_2 n)^4$. In addition, if $q_1^{N_1-1} = n+1$, then $N_1 = \frac{\log_2(n+1)}{4 \log_2({1065} \log_2(n))} + 1$. Then, $r_1 \leq \frac{7}{3} \cdot N_1$, and so
\begin{align}
\log_2 q_1^{r_1} \leq& {\frac{7}{3} \log_2 (n+1) + \frac{28}{3} \log_2 (1065 \log_2(n))} \label{eq:qr1} \\
=&\frac{7}{3} \log_2(n+1) + \cO(\log_2 \log_2 n). \nonumber
\end{align}
Assuming $Q=s+1= {1065} \log_2 (n) + 1$, then we {approximate} $q_2 = {2130} \log_2(n) + 2$. In addition if $q_2^{N_2} = n+2$, then $N_2 = \frac{\log_2(n+2)}{\log_2( {2130} \log_2(n) + 2)}$. Since $r_2 \leq 1 + N_2$, we have
\begin{align}
\log q_2^{r_2} &\leq {\log_2(n+2) + \log_2(2130 \log_2(n) + 2)} \label{eq:qr2} \\ 
&=\log_2(n+2) + \cO(\log_2 \log_2(n)). \nonumber
\end{align}
Thus, 
\begin{align*}
\log_2 |\cC^{(2)}(n,& {\bfa_{0000}}, {\bfa_{1111}}, \bfa_{110011}, \bfa_{11011}, \bfc, b, s)|  \geq \\
&\log_2 |\cC_{T2}(n,s)| - 8 \log n  - \cO(\log_2 \log_2 n).
\end{align*}
In the next section, we show that when {$s \geq 1065 \log_2(n)$}, $\log_2 |\cC_{T2}(n,s)| \geq n - \cO(1)$ and so there exists a code which meets our lower bound.

We now prove that the code $\cC^{(2)}(n, {\bfa_{0000}}, {\bfa_{1111}}, \bfa_{110011}, \bfa_{11011},$ $\bfc, b, s)$ can correct two deletions by considering the same scenarios as in the previous section. With a slight abuse of notation, in this section $\cC(n)$ will denote  $\cC^{(2)}(n, {\bfa_{0000}}, {\bfa_{1111}}, \bfa_{110011},$ $\bfa_{11011}, \bfc, b, s)$ and NOT $\cC(n,{\bfa_{0000}}, {\bfa_{1111}}, \bfa_{110011}, \bfa_{11011}, \bfc, s)$ from the previous section.

Analagous to the previous section, we begin with the following claim and throughout we assume $\bfx \in \cC(n)$.

\begin{claim}\label{cl:if1101110112} Suppose a zero is deleted from a run of length one in $\bfx$ and the deletion causes
\begin{enumerate}
\item $110011$ to be created/destroyed,
\item $1111$ to be created,
\end{enumerate}
then the substring $11011$ is preserved from $\bfx$ to $\bfy$, and $d_H \left(F_{h_s^{(R)}}(\bfx,11011),F_{h_s^{(R)}}(\bfy,11011) \right)=1$. \end{claim}
\begin{IEEEproof} The deletion of a zero from a run of length $1$ {clearly cannot destroy a $110011$ substring}. The only other case to consider is when a substring $110011$ is created and also the substring $1111$ is created. Notice that this is only possible when the first zero is deleted from the substring $111010011$ or if the last zero is deleted from the substring $110010111$. Notice that under either setup, the substring $11011$ is preserved. In addition ${\tau}_{\geq 2}(111010011) = (3,2)$ and ${\tau}_{\geq 2}(11110011) = (4,2)$ so that $d_H \left(F_{h_s^{(R)}}(\bfx,11011),F_{h_s^{(R)}}(\bfy,11011) \right)=1$ (notice ${\tau}_{\geq 2}(110010111)=(2,3)$ and ${\tau}_{\geq 2}(11001111)=(2,4)$ ). \end{IEEEproof}

\begin{claim}\label{cl:ifr2r1} Suppose that $\bfy$ is the result of deleting a zero in $\bfx$ from a run of length $1$ where the zero is adjacent to runs of length $1$ and length $\ell$ where $\ell \geq 3$. Then, given ${\tau}_{\geq 2}(\bfx)$, $\sum_{i \text{\ odd}} {\tau}_1(\bfx)_i \bmod {(}s+1 {)}$, and $\bfy$, it is possible to determine ${\tau}_1(\bfx)$.
\end{claim}
\begin{IEEEproof} Since $\bfy$ is the result of deleting a zero from a run of length $1$ where the zero is adjacent to runs of length $1$ and $\ell \geq 3$, it follows that ${\tau}_{\geq 2}(\bfy)$ can be obtained by substituting a symbol in ${\tau}_{\geq 2}(\bfy)$ which has value $\ell$ with another symbol which has value $\ell+1$. Since $\ell+1 \geq 3$, it follows that $d_H({\tau}_{\geq 2}(\bfy), {\tau}_{\geq 2}(\bfx)) = 1$ and so we can determine the location of the symbol in ${\tau}_{\geq 2}(\bfy)$ that was altered as a result of the deletion to $\bfy$. To obtain ${\tau}_1(\bfx)$ from ${\tau}_1(\bfy)$, ${\tau}_{\geq 2}(\bfx)$, ${\tau_{\geq 2}(\bfy)}$, we replace the symbol, say $a$, that has value $\ell+1$ in ${\tau}_1(\bfy)$ which corresponds to the same symbol in ${}{\tau}_{\geq 2}(x)$ (that was affected by the deletion of the zero in $\bfx$) and we replace the symbol $a$ in ${}{\tau}_1(\bfy)$ with 2 adjacent symbols $1$ and $\ell$. Given $\sum_{i \text{\ odd}} {}{\tau}_1(\bfx)_i  \bmod {}{(}s+1{}{)}$, we can determine whether the symbol $1$ should be inserted before the symbol $\ell$ or whether $\ell$ comes before the symbol $1$. Thus, we can recover ${}{\tau}_1(\bfx)$ as stated in the claim. \end{IEEEproof} \vspace{1.5ex}

We have the following lemmas which mirror the logic from the previous section. The first lemma follows immediately using the same logic as in the proof of Lemma~\ref{lem:2o}.

\begin{lemma} Suppose $N_1(\bfx) - N_1(\bfy) \bmod 7 = 2$. Then, $\bfx$ can be recovered from $\bfy$. \end{lemma}

The next lemma requires a little more work.

\begin{lemma}\label{lem:zeros2} Suppose $N_0(\bfx) - N_0(\bfy) \bmod 7 = 2$. Then, $\bfx$ can be recovered from $\bfy$. \end{lemma}
\begin{IEEEproof} Similar to the proof of Lemma~\ref{lem:zeros}, we focus on the case where both ${}{N_{1111}}(\bfx) - {}{N_{1111}}(\bfy) \not \equiv 0 \bmod 7$ and ${}{N_{0000}}(\bfx) - {}{N_{0000}}(\bfy) \not \equiv 0 \bmod 7$, since if at most one of these two conditions hold then we can determine $\bfx$ from $\bfy$ given $F_{h_s}(\bfx, 0000) \in \cC_2(n,q,{}{\bfa_{0000}}), F_{h_s}(\bfx, 1111)  \in \cC_2(n,q, {}{\bfa_{1111}})$. 

Since ${}{N_{1111}}(\bfx) - {}{N_{1111}}(\bfy) \not \equiv 0 \bmod 7$ and ${}{N_{0000}}(\bfx) - {}{N_{0000}}(\bfy) \not \equiv 0 \bmod 7$ hold, it follows that $\bfy$ is the result of deleting a zero from a run of length $1$ and another zero from a run of length at least $4$. Notice that the deletion of the zero from a run of length $4$ cannot create/destroy the substrings $110011$, $1111$.  Thus, if the substring $110011$ is not preserved from $\bfx$ to $\bfy$, it is a result of the deletion of the zero from a run of length $1$. According to Claim~\ref{cl:if1101110112}, under this setup, $d_H \left(F_{h_s^{(R)}}(\bfx,11011),F_{h_s^{(R)}}(\bfy,11011) \right)=1$. Thus, we can determine ${}{\tau}_{\geq 2}(\bfx)$ from $F_{h^{(R)}_{s}}(\bfx, 11011) \in \cC_1(n,q_2,\bfa_{11011})$. According to Claim~\ref{cl:ifr2r1}, we can then determine ${}{\tau}_1(\bfx)$ so that we can correct the deletion of the zero from a run of length $1$. The remaining deletion (of a zero from a run of length $\geq 4$) can be corrected using the constraint $F_{h_s}(\bfx, 1111)  \in \cC_2(n,q_1,\bfa_2)$.

Thus, we can determine $\bfx$ from $\bfy$ as follows. Suppose ${}{N_{1111}}(\bfx) - {}{N_{1111}}(\bfy) \not \equiv 0 \bmod 7$, ${}{N_{0000}}(\bfx) - {}{N_{0000}}(\bfy) \not \equiv 0 \bmod 7$, and $ N_{110011}(\bfx) \not \equiv  N_{110011}(\bfy) \bmod 7$. Then, $\bfx$ can be recovered as described in the previous paragraph using the constraints $ F_{h^{(R)}_{s}}(\bfx, 11011) \in \cC_1(n,q_2,\bfa_{11011}), F_{h_s}(\bfx, 1111)  \in \cC_2(n,q_1,{}{\bfa_{1111}})$. Otherwise, if ${}{N_{1111}}(\bfx) - {}{N_{1111}}(\bfy) \not \equiv 0 \bmod 7$, ${}{N_{0000}}(\bfx) - {}{N_{0000}}(\bfy) \not \equiv 0 \bmod 7$, and $ N_{110011}(\bfx)  \equiv  N_{110011}(\bfy) \bmod 7$, $\bfx$ can be recovered from $\bfy$ using $F_{h_s}(\bfx, 110011) \in \cC_2(n,q_1,\bfa_{110011})$. \end{IEEEproof}\vspace{1.5ex}

The next lemma can be proven in the same manner as Lemma~\ref{lem:11}.

\begin{lemma} Suppose ${}{N_{0000}}(\bfx) - {}{N_{0000}}(\bfy) \bmod 7 = 1$, and ${}{N_{1111}}(\bfx) - {}{N_{1111}}(\bfy) \bmod 7 = 1$. Then, $\bfx$ can be recovered from $\bfy$. \end{lemma} 

The proofs of the next two lemmas are the same as Lemma~\ref{lem:10} and Lemma~\ref{lem:01}.

\begin{lemma} Suppose ${}{N_{0000}}(\bfx) - {}{N_{0000}}(\bfy) \bmod 7 = 1$ and ${}{N_{1111}}(\bfx) - {}{N_{1111}}(\bfy) \bmod 7 = 0$. Then, $\bfx$ can be recovered from $\bfy$.
\end{lemma}

\begin{lemma} Suppose ${}{N_{1111}}(\bfx) - {}{N_{1111}}(\bfy) \bmod 7 = 1$ and ${}{N_{0000}}(\bfx) - {}{N_{0000}}(\bfy) \bmod 7 = 0$. Then, $\bfx$ can be recovered from $\bfy$.
\end{lemma}

Next, we consider the case where either ${}{N_{0000}}(\bfy) > {}{N_{0000}}(\bfx)$ or ${}{N_{1111}}(\bfy) > {}{N_{1111}}(\bfx)$. The result can be proven using ideas similar to Lemma~\ref{lem:0111} and Lemma~\ref{lem:zeros2}. The proof can be found in Appendix~\ref{app:01plus}.

\begin{lemma}\label{lem:01plus} Suppose ${}{N_{0000}}(\bfy) - {}{N_{0000}}(\bfx) \bmod 7 \in \{1,2,3\}$ or ${}{N_{1111}}(\bfy) - {}{N_{1111}}(\bfx) \bmod 7 \in \{1,2,3\}$. Then, $\bfx$ can be recovered from $\bfy$.
\end{lemma}

The next lemma follows from Lemma~\ref{lem:finallem}. 

\begin{lemma}Suppose ${}{N_{1111}}(\bfx) - {}{N_{1111}}(\bfy) \bmod 7 = 0$, and ${}{N_{0000}}(\bfx) - {}{N_{0000}}(\bfy) \bmod 7 = 0$. Then, $\bfx$ can be recovered from $\bfy$.
\end{lemma}

\section{Constraint Redundancy}\label{sec:constraint}
The purpose of this section is to show that there is no asymptotic rate loss incurred by starting with our constrained sequence space where there are no more than $s$ symbols between consecutive appearances of ${\bf v}_1 = 0000$, ${\bf v}_2 = 1111$, ${\bf v}_3 = 11011$, ${\bf v}_4 = 110011$, $\bfv_5 = 1$ and $\bfv_6 = 0$. Our goal will be to show that the probability a sequence of length $n$ satisfies these constraints converges to {{} 1} for sufficiently large $n$. This implies that the redundancy incurred is indeed a constant number of bits.

{{} We first sketch our approach. We divide the sequence into subwords of length $s/2$. We then lower bound the probability of the event that ${\bf v}_i$ appears in each of these length $s/2$ subwords, so that indeed there cannot be more than $s$ symbols between any two appearances of ${\bf v}_i$. 

Let $A_{N,{\bf v}}$ be the probability that a sequence of length $N$ selected uniformly at random from $\{0,1\}^N$ contains the sequence $\bf v$. We show that 
\begin{lemma}
For $1 \leq i \leq 6$, 
\[A_{N,{\bf v}_i} \geq 1-\exp\left( -\frac{N}{|{\bf v}_i|}2^{-|{\bf v}_i|-1} \right).\]
\label{lem:useones}
\end{lemma} 

%
%
%

This enables us to lower bound the probability that ${\bf v}_i$ is found in each length $n/(s/2)$ subword. This probability is $(A_{s/2,{\bf v}_i})^{n/(s/2)}$, and we have that 
\begin{align*}
(A_{s/2, {\bf v}_i})^{n/(s/2)} \geq \left( 1-\exp\left( -\frac{s/2}{|{\bf v}_i|}2^{-|{\bf v}_i|-1} \right) \right)^{n/(s/2)}.
\end{align*}
It remains to show that for $s = O(\log_2(n))$ the right-hand side above goes to 1:
\begin{lemma}
\label{lem:finish}
There exists a constant $c$ such that if $s = c \log_2(n)$, then $(A_{s/2, {\bf v}_i})^{n/(s/2)} \rightarrow 1$. 
\end{lemma}
We are nearly done. Since the probabilities $A_{N,{\bf v}_1},A_{N,{\bf v}_4},A_{N,{\bf v}_5},$ and $A_{N,{\bf v}_6}$ all go to 1 as $n \rightarrow \infty$, a union bound argument shows that the probability they all hold simultaneously also converges to 1, completing the proof.

Next, we prove the two lemmas. 

\begin{IEEEproof}
\emph{Lemma~\ref{lem:useones}.} For notational convenience, we write $\ell = |{\bf v}_i|$. We break up the string of length $N$ into $N/\ell$ substrings of length $\ell$. The probability $A_{N,\ell}$ is lower bounded by the probability of at least one of the $N/\ell$ substrings being ${\bf v}_i$. Consider the Bernoulli random variables $X_j$ ($1 \leq j \leq N/\ell$) that have value 1 if the $j$th substring is ${\bf v}_i$ (with probability $2^{-\ell}$) and 0 otherwise. 

Now we apply the multiplicative version of the Chernoff bound on $X = \sum_{j=1}^{N/\ell} X_j$. This bound states that $Pr(X \leq (1-\delta)E[X]) \leq \exp(-\delta^2 E[X]/2)$. Note that the mean of $X$ is $E[X] = (N/\ell) 2^{-\ell}$. Taking $\delta \rightarrow 1$, we have  
\[Pr( X = 0)  \leq \exp\left(- \frac{N 2^{-\ell-1}}{\ell} \right).\]
Thus, the probability of at least one appearance of ${\bf v}_i$ is lower bounded by
\[1-\exp\left(- \frac{N 2^{-\ell-1}}{\ell} \right),\]
as desired.
\end{IEEEproof}

\begin{IEEEproof}
\emph{Lemma~\ref{lem:finish}.} Again we set $\ell = |{\bf v}_i|$. The substring length $N$ is now $s/2 = c\log_2(n)/2$. Set 
\begin{align*}
X_n &= \left[ 1-\exp \left( -\frac{(c\log_2 n)/2(2^{-\ell-1})}{\ell} \right)    \right]^{\frac{2n}{c\log_2 n}} \\
&=\left[ 1-\exp \left(-\frac{(c\log_2 n)(2^{-\ell-2})}{\ell} \right)    \right]^{\frac{2n}{c\log_2 n}}
\end{align*}
so that according to Lemma~\ref{lem:useones}, $(A_{s/2, {\bf v}_i})^{n/(s/2)} \geq X_n$.    
We use the Taylor series expansion for $\log(1+x)$ to write
\begin{align*}
\log X_n &= \frac{2n}{c \log_2 n} \log \left[ 1-\exp \left(-\frac{(c\log_2 n)(2^{-\ell-2})}{\ell} \right)    \right] \\
&= \frac{2n}{c \log_2 n} \log \left(1- n^{-\frac{2^{-\ell-2}c}{\ell \log 2}} \right)\\
&=  \frac{2n}{c \log_2 n} \left( n^{-\frac{2^{-\ell-2}c}{\ell \log 2}} - O\left( \left( n^{-\frac{2^{-\ell-2}c}{\ell \log 2}}\right)^2\right)\right).
\end{align*}
The last step follows from the expansion of $\log(1+x)$. Now, if 
\[\frac{2^{-\ell-2}c}{\ell \log 2} > 1,\]
the terms inside the parentheses will dominate the $2n$ factor outside and we will have $\lim_{n \rightarrow \infty} \log X_n \rightarrow 0$, so that $X_n \rightarrow 1$, as desired. For our constraints, the largest length $\ell$ of a constraint string ${\bf v}_i$ is $\ell=6$. We just need $c$ such that $c > \ell 2^{\ell+2} \log 2$; therefore we can take any $c \geq 1065$.
\end{IEEEproof}
We note that the value of the constant $c$ can be reduced by bounding the probability $A_{s/2,{\bf v}_i}$ more tightly; this is possible through a more involved argument based on recursive bounds.

}

\section{Conclusion}\label{conc}

In this work, we provided a construction for a code capable of correcting two deletions that improved upon existing art in terms of the number of redundant bits. {}{Our approach relied on requiring that for any codeword $\bfx$ we have that $F_{h_s}(\bfx,0000)$, $F_{h_s}(\bfx, 1111)$, $F_{h_s}(\bfx, 11011)$, and $F_{h_s}(\bfx, 110011)$ each belong to a code with minimum Hamming distance $5$. A natural extension of this work would be to consider the construction of codes capable of correcting three or more deletions. As a starting point, it can be shown that if $\bfy = \cD_3(\bfx)$ is the result of $3$ deletions occurring to $\bfx$, then at least one of the following substrings is preserved from $\bfx$ to $\bfy$:}

{}{$$ \Big \{ (1,1,1,1,1,1), (0,0,0,0,0,0), (1,1,1,0,1,1,1), (1,1,1,0,0,1,1,1), (1,1,1,0,0,0,1,1,1) \Big \}. $$}{}{Therefore, using the same ideas as in this work, we would require $F_{h_s}(\bfx,000000)$, $F_{h_s}(\bfx, 111111)$, $F_{h_s}(\bfx, 1110111)$, $F_{h_s}(\bfx, 11100111)$, and $F_{h_s}(\bfx, 111000111)$ each belong to a code with minimum Hamming distance $7$. Note that this would require at least $(3 \log n) \cdot 4 = 12 \log n$ bits of redundancy. The challenge in establishing this result for $t \geq 3$ is to handle the extensive amount of casework which would be required if the techniques in this paper were adopted, and it is not immediately clear how to proceed for general $t$.}

{}{Another area of future work involves devising efficient encoding algorithms. This will likely require a two step process, one for generating the initial sequences that satisfy our constraints, and the other for the error-correction properties required for the constructions. One angle of attack would be to use a similar approach to error-correcting constrained codes, which concatenate the error-correction encoder with the constrained encoder; however, finding an efficient technique is likely to be challenging.}

\begin{appendices}

\section{Proof of Lemma~\ref{lem:01plus}}\label{app:01plus}

\noindent \textbf{Lemma 15.} Suppose ${}{N_{0000}}(\bfy) - {}{N_{0000}}(\bfx) \bmod 7 \in \{1,2,3\}$ or ${}{N_{1111}}(\bfy) - {}{N_{1111}}(\bfx) \bmod 7 \in \{1,2,3\}$. Then, $\bfx$ can be recovered from $\bfy$.
\begin{IEEEproof} We consider the case where  ${}{N_{0000}}(\bfy) - {}{N_{0000}}(\bfx) \bmod 7 \in \{1,2,3\}$ and ${}{N_{1111}}(\bfy) - {}{N_{1111}}(\bfx) \bmod 7 \in \{1,2,3\}$ since otherwise $\bfx$ can be recovered from $\bfy$ using the same ideas as in the proof of Lemma~\ref{lem:0111}.  The deletion of a one from a run of length $1$ cannot create/destroy a $110011$ substring or a $11011$ substring since the $1$ needs to be adjacent to two runs of lengths $\ell_1, \ell_2$ where $\ell_1 + \ell_2 \geq 4$ since at least one $0000$ substring is created from $\bfx$ to $\bfy$. Furthermore, the deletion of the one from a run of length $1$ clearly cannot create a $1111$ substring from $\bfx$ to $\bfy$. Therefore, the deletion of the zero from a run of length $1$ creates a $1111$ substring from $\bfx$ to $\bfy$. If, in addition, the deletion of the zero does also does not preserve the $110011$ substring from $\bfx$ to $\bfy$, then according to Claim~\ref{cl:if1101110112},  the substring $11011$ is preserved from $\bfx$ to $\bfy$ and $d_H \left(F_{h_s^{(R)}}(\bfx,11011),F_{h_s^{(R)}}(\bfy,11011) \right)=1$. Using the same logic as in the proof of Lemma~\ref{lem:zeros2}, according to Claim~\ref{cl:ifr2r1} we can correct the deletion of a zero from a run of length $1$ using the constraints $F_{h^{(R)}_{s}}(\bfx, 11011) \in \cC_1(n,q_2,\bfa_{11011})$ and $\sum_{i \text{\ odd}} {}{\tau}_1(\bfx)_i = b \bmod {}{(}s + 1{}{)}$ and the deletion of the one from a run of length $\geq 4$ can be corrected with the constraint $F_{h_s}(\bfx, 0000)  \in \cC_2(n,q_1,{}{\bfa_{0000}})$. Thus, the decoding in this case is the same as described in the last paragraph of Lemma~\ref{lem:zeros2}. 
\end{IEEEproof}

\end{appendices}

\end{document}